\documentclass[9pt]{extarticle}
\usepackage{spconf,amsmath,graphicx,hyperref}
\usepackage{amsmath}
\usepackage{graphicx}
\usepackage{rotating}
\usepackage{pdflscape}
\usepackage{multirow}
\usepackage{tikz}
\usepackage[utf8]{inputenc} 
\usepackage[T1]{fontenc}    
\usepackage{hyperref}       
\usepackage{url}            
\usepackage{booktabs}       
\usepackage{amsfonts}       
\usepackage{nicefrac}       
\usepackage{microtype}      
\usepackage{xcolor}         

\usepackage{booktabs}
\usepackage{tabularx}
\usepackage{makecell}
\usepackage{array}
\usepackage{threeparttable}  

\newcolumntype{Y}{>{\raggedright\arraybackslash}X}   
\newcolumntype{C}{>{\centering\arraybackslash}p{0.18\textwidth}}
\newcolumntype{F}{>{\raggedright\arraybackslash}p{0.25\textwidth}}

\title{FINSENTLLM: MULTI-LLM AND STRUCTURED SEMANTIC SIGNALS FOR ENHANCED FINANCIAL SENTIMENT FORECASTING}

\name{
\makecell[c]{Zijian Zhang$^{1\dag\ast}$\thanks{$^{\ast}$Corresponding author: Zijian Zhang (\texttt{zzjharry@alumni.upenn.edu})}, Rong Fu$^{2\dag}$\thanks{$^{\dag}$ These authors contribute to this work equally.}, Yangfan He$^{3}$, Xinze Shen$^{4}$, Yanlong Wang$^{5}$ \\ Xiaojing Du$^{6}$, Haochen You$^{7}$, Jiazhao Shi$^{8}$, Simon Fong$^{2}$
}
}

\address{$^{1}$University of Pennsylvania, $^{2}$University of Macau, $^{3}$University of Minnesota, $^{4}$University of Birmingham \\$^{5}$Tsinghua University, $^{6}$University of South Australia, $^{7}$Columbia University, $^{8}$New York University}

\begin{document}
%
\maketitle

\begingroup
\renewcommand\thefootnote{}
\footnotetext{%
\textcopyright~2025 IEEE. Personal use of this material is permitted. Permission from IEEE must be obtained for all other uses, in any current or future media, including reprinting/republishing this material for advertising or promotional purposes, creating new collective works, for resale or redistribution to servers or lists, or reuse of any copyrighted component of this work in other works.}
\addtocounter{footnote}{-1}
\endgroup

\begin{abstract}
Financial sentiment analysis (FSA) has attracted significant attention, and recent studies increasingly explore large language models (LLMs) for this field. Yet most work evaluates only classification metrics, leaving unclear whether sentiment signals align with market behavior. We propose \textsc{FinSentLLM}, a lightweight multi-LLM framework that integrates an expert panel of sentiment forecasting LLMs, and structured semantic financial signals via a compact meta-classifier. This design captures expert complementarity, semantic reasoning signal, and agreement/divergence patterns without costly retraining, yielding consistent 3–6\% gains over strong baselines in accuracy and F1-score on the Financial PhraseBank dataset. In addition, we also provide econometric evidence that financial sentiment and stock markets exhibit statistically significant long-run comovement, applying Dynamic Conditional Correlation GARCH (DCC–GARCH) and the Johansen cointegration test to daily sentiment scores computed from the FNSPID dataset and major stock indices. Together, these results demonstrate that \textsc{FinSentLLM} delivers superior forecasting accuracy for financial sentiment and further establish that sentiment signals are robustly linked to long-run equity market dynamics.
\end{abstract}

\section{Introduction}
\label{sec:intro}

Financial sentiment analysis (FSA) has been a longstanding research area, with early studies relying on lexicon-based~\cite{oliveira2016stock} and traditional machine learning methods~\cite{atzeni2017fine}. In recent years, FinBERT~\cite{araci2019finbert} and other fine-tuned pre-trained language models (PLM) on financial corpora \cite{du2024financial}, achieving strong results on benchmarks such as the Financial PhraseBank (FPB)~\cite{Malo2014GoodDO} and FiQA~\cite{shah2022flue}. More recently, large language models (LLMs) such as GPT-4 and LLaMA-7B have been explored in zero- and few-shot FSA, often rivaling supervised baselines~\cite{li2023chatgpt, zhang2023instruct}. However, these efforts largely emphasize intrinsic NLP metrics (e.g., Macro-F1), leaving open the question of whether sentiment signals derived from LLMs align with financial market behavior.

A second line of work has investigated the predictive value of sentiment for markets, constructing indices from news or social media and incorporating them into time-series models such as LSTMs and Informer.~\cite{hossain2024finbert, dong2025deep} While these studies report modest gains in forecasting accuracy or portfolio performance, sentiment is often treated as an auxiliary feature, with limited analysis of its structural relation to asset returns.

A third body of research applies econometric methods to sentiment–market interactions. Examples include Transfer Entropy with EGARCH for nonlinear information flow~\cite{mendoza2022twitter} and DCC–GARCH frameworks for dynamic correlation~\cite{pillada2023empirical}. While these studies show that sentiment can be linked to markets, their sentiment measures are typically lexicon-based or derived from traditional classifiers rather than modern LLMs.

Our work differs from all three streams in several key ways. 
First, we propose \textsc{FinSentLLM}, a lightweight framework that integrates domain-specific and general-purpose LLMs with structured financial semantics via a compact meta-classifier. Without any additional fine-tuning, \textsc{FinSentLLM} substantially outperforms strong domain-specific baselines (e.g., FinBERT) on the FPB dataset, achieving consistent gains in accuracy and Macro-F1. 
Second, our framework introduces a novel architecture that combines an expert panel of sentiment LLMs with high-precision financial semantic signals, enabling complementary reasoning beyond any single model. 
Third, we provide econometric evidence of financial relevance. By applying DCC–GARCH and Johansen cointegration to FinBERT-derived sentiment scores and major stock indices, we show that sentiment and equity markets exhibit statistically significant long-run comovement. 

To our knowledge, this is the first study that both demonstrates the predictive advantages of multi-LLM and semantic financial signals integration for financial sentiment forecasting and establishes, through econometric analysis, that the financial sentiment signals align with persistent market linkages.

\begin{figure*}[t]
    \centering
    \includegraphics[width=\linewidth]{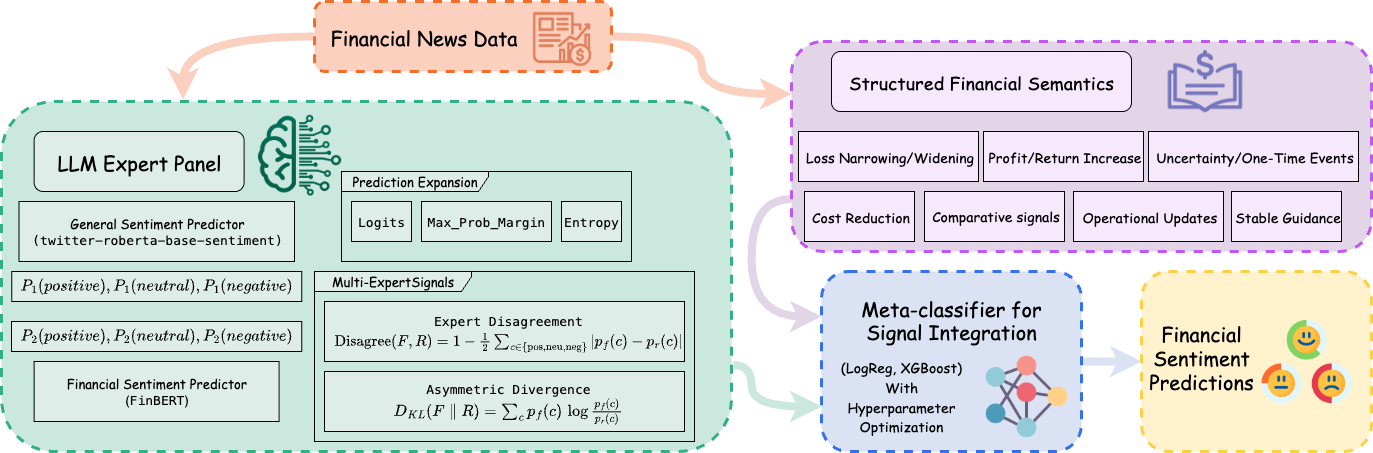}
    \caption{Overall architecture of the proposed \textsc{FinSentLLM} framework. The design integrates probability-derived features, structured financial semantics, and multi-LLM expert signals, which are unified by a meta-classifier for signal integration to enhance financial sentiment forecasting.}
    \label{fig:fsa_framework}
\end{figure*}

\section{Methodology}

An overview of the proposed framework, \textsc{FinSentLLM}, is shown in Fig.~\ref{fig:fsa_framework}. \textsc{FinSentLLM} is a training-free integration framework for financial sentiment analysis that is both efficient and extensible. Concretely, we employ two LLM expert predictors: (i) FinBERT, a BERT-based model pre-trained on financial corpora and fine-tuned for finance sentiment, and (ii) Twitter-RoBERTa-base-sentiment~\cite{barbieri-etal-2020-tweeteval} (abbrev.\ RoBERTa-sentiment), a RoBERTa model trained on 58M tweets and fine-tuned on TweetEval. Beyond per-model posteriors, we explicitly aggregate these experts’ predictions to build Multi-LLM expert signals, including confidence asymmetry, agreement, and directional divergence, so that inter-expert complementarities are directly captured rather than implicitly assumed. In parallel, we inject structured financial semantics derived from text patterns (e.g., profit increase, cost reduction, contract/financing announcements). All signals are unified by a meta-classifier for signal integration (logistic regression or XGBoost) that outputs the final decision.

This design leverages diverse strengths while mitigating weaknesses: FinBERT offers domain precision but may inherit finance-specific biases; RoBERTa-sentiment brings broader linguistic coverage. Multi-LLM expert signals summarize how the experts relate to each other on each instance (e.g., strong agreement vs.\ informative disagreement), and structured semantic flags encode high-precision financial reasoning cues. The meta-classifier combines these heterogeneous inputs without any large-scale fine-tuning, yielding a simple and efficient stacking-style integration.

\subsection{Representation Design for Financial Sentiment Signals}

Starting from the posterior distributions of FinBERT and RoBERTa-sentiment over \textit{positive}, \textit{neutral}, and \textit{negative} classes, we construct three groups of financial sentiment signals:

\textbf{(i) Probability-derived features}: logits, maximum probability, margin, and entropy;  

\textbf{(ii) Structured financial semantics}: binary flags triggered by domain-specific expressions such as comparative patterns, profit/return increase, cost reduction, or contract announcements;  

\textbf{(iii) Multi-LLM expert signals}: model-specific confidence, agreement, and divergence between FinBERT and RoBERTa-sentiment.  

Table~\ref{tab:feature_inventory_4col} summarizes these financial sentiment signals with definitions and rationale. Together, they enable the meta-classifier to detect agreement, confidence imbalance, or contradictions between experts, yielding decisions more robust to systematic biases.

\begin{table*}[t]
\setlength{\tabcolsep}{3pt}               
\caption{Summary of representation design for financial sentiment signals.}
\vspace{6pt}
\renewcommand{\arraystretch}{1.05}
\footnotesize
\centering
\begin{threeparttable}
\resizebox{\linewidth}{!}{%
\begin{tabular}{>{\centering\arraybackslash}m{0.16\linewidth}%
                >{\centering\arraybackslash}m{0.20\linewidth}%
                >{\raggedright\arraybackslash}m{0.32\linewidth}%
                >{\raggedright\arraybackslash}m{0.28\linewidth}}
\toprule
\multicolumn{1}{c}{\textbf{Direction}} & 
\multicolumn{1}{c}{\textbf{Signal}} & 
\multicolumn{1}{c}{\textbf{Definition / Formula}} &
\multicolumn{1}{c}{\textbf{Rationale}} \\
\midrule

\multirow{4}{*}{\centering\arraybackslash\makecell{\\\textbf{Probability-Derived}\\\textbf{Features}}}
& \textbf{Logits} & $\mathrm{logit}(p_c)=\log \tfrac{p_c}{1-p_c}$ & Linearizes probabilities; highlights boundary cases. \\
\cmidrule(lr){2-4}
& \textbf{Max prob} & $\max(p_{\text{pos}},p_{\text{neu}},p_{\text{neg}})$ & Captures model’s confidence. \\
\cmidrule(lr){2-4}
& \textbf{Margin} & $\mathrm{top1}(p)-\mathrm{top2}(p)$ & Measures decisiveness of prediction. \\
\cmidrule(lr){2-4}
& \textbf{Entropy} & $H(p)=-\sum_{c} p_c \log p_c$ & Quantifies overall uncertainty. \\
\midrule

\multirow{8}{=}{\centering\arraybackslash\makecell{\\\\\\\\\textbf{Structured Financial}\\\textbf{Semantics}}}
& \textbf{Comparative patterns} & Pattern: ``compared to'', ``from–to'' & Detects relative change expressions. \\
\cmidrule(lr){2-4}
& \textbf{Loss $\downarrow$/$\uparrow$} & Pattern: ``loss narrowed'' ($\downarrow$), ``loss widened'' ($\uparrow$) & Distinguishes improvement vs.\ deterioration. \\
\cmidrule(lr){2-4}
& \textbf{Profit/return $\uparrow$} & Keywords: ``profit'', ``ROE'', ``ROI'' + upward verbs (``rose'', ``grew'') & Flags positive performance indicators. \\
\cmidrule(lr){2-4}
& \textbf{Cost $\downarrow$} & Keywords: ``fell'', ``lower'', ``reduced'' & Identifies cost reduction signals. \\
\cmidrule(lr){2-4}
& \textbf{Contract/financing} & Keywords: ``agreement'', ``bond issuance'', ... & Neutral corporate actions often misclassified. \\
\cmidrule(lr){2-4}
& \textbf{Uncertainty/one-off} & Keywords: ``uncertain'', ``cannot be determined'', ... & Captures ambiguous or one-time events. \\
\cmidrule(lr){2-4}
& \textbf{Stable guidance} & Pattern: ``expects to remain'', ``reiterated forecast'' & Marks neutral forward-looking statements. \\
\cmidrule(lr){2-4}
& \textbf{Operational updates} & Keywords: ``restructuring'', ``ban'', ... & Procedural, non-sentiment updates. \\
\midrule
\multirow{3}{*}{\centering\arraybackslash\makecell{\textbf{Multi-LLM Expert}\\\textbf{Signals}}}
& \textbf{Expert disagreement (L1)} & $1-\tfrac{1}{2}\sum_{c}|p_{f}(c)-p_{r}(c)|\in[0,1]$ & Measures similarity of experts’ distributions. \\
\cmidrule(lr){2-4}
& \textbf{Kullback--Leibler (KL) divergence} & $D_{KL}(F \parallel R) = \sum_{c} p_{f}(c)\,\log\frac{p_{f}(c)}{p_{r}(c)}$ & Captures directional disagreement among experts. \\
\bottomrule
\end{tabular}
}
\begin{tablenotes}[flushleft]
\footnotesize
\item \textbf{Notes:} $p_c$: posterior probability of class $c\in\{\text{pos},\text{neu},\text{neg}\}$. 
$p_f(c),p_r(c)$: probabilities from FinBERT (F) and RoBERTa-sentiment (R). 
$\mathrm{top1}(p),\mathrm{top2}(p)$: highest/second-highest probabilities. 
$H(p)$: Shannon entropy. 
$D_{KL}(\cdot\parallel\cdot)$: Kullback--Leibler divergence~\cite{cover1999elements}.
\end{tablenotes}
\end{threeparttable}
\label{tab:feature_inventory_4col}
\vspace{-2mm}
\end{table*}

\subsection{Meta-classifier for Signal Integration}

The derived representations are fed into a lightweight supervised meta-classifier, which integrates heterogeneous signals from FinBERT, RoBERTa-sentiment, and structured financial semantics. We experiment with logistic regression and gradient boosted trees (XGBoost). The latter is particularly effective at modeling non-linear interactions, for example when a semantic flag alters the interpretation of probability-derived features.

\vspace{3pt}
\noindent\textsc{FinSentLLM} balances complementarity (domain-specific vs. general-domain experts), domain knowledge injection (via semantic flags), and efficiency (lightweight integration without large-scale fine-tuning), while remaining extensible to new LLM experts.

\section{Experimental Setup}

We conduct comprehensive experiments to evaluate the effectiveness of our proposed \textsc{FinSentLLM} framework against strong baselines. To this end, we describe the benchmark datasets used for sentiment classification, the large-scale news corpus employed for market-linkage analysis, the preprocessing steps applied to ensure consistency, and the baseline models and configurations used for comparison.

\subsection{Datasets}

\textbf{Financial PhraseBank (FPB)}.  
This benchmark dataset consists of 14,780 unique financial news headlines, each annotated with sentiment labels (\textit{positive}, \textit{neutral}, or \textit{negative}) by domain experts. To account for annotation uncertainty, FPB provides four subsets corresponding to different levels of annotator agreement: 50\%, 66\%, 75\%, and 100\%. We use FPB as the primary dataset for evaluating classification performance against baselines.

\vspace{3pt}
\noindent\textbf{FNSPID}~\cite{dong2024fnspid}.  
This large-scale dataset contains 15,698,563 financial news articles collected from 1999–2023 across sources including Bloomberg, Reuters, Benzinga, and Lenta. In this work, we focus on the 2018–2019 period, covering 501 trading days and 679,795 news items. Stock market indices (log of closing prices) used for linkage analysis are obtained from \texttt{yfinance}.

\subsection{Data Preprocessing}

For FPB, preprocessing is minimal and limited to deduplication of identical headlines.  
For FNSPID, we first apply sentence-level sentiment scoring with FinBERT. 
Daily sentiment scores are then constructed as 
$s_t = p_{\text{pos}} - p_{\text{neg}} - 0.1 \times p_{\text{neu}}$, 
and standardized using z-scores. Finally, market index prices are aligned with sentiment scores at the daily frequency to enable econometric analysis.

\subsection{Baselines and Configurations}

We benchmark \textsc{FinSentLLM} against two categories of baselines. 
First, we consider domain-specific sentiment experts, including FinBERT and RoBERTa-sentiment, 
which correspond to the constituent predictors within our framework. Second, we evaluate general-purpose large language models, specifically OpenAI GPT-4o mini
and GPT-5, which provide competitive zero-shot sentiment predictions.~\cite{radford2018improving}

For the meta-classifier for signal integration in \textsc{FinSentLLM}, we employ logistic regression and XGBoost classifiers. The XGBoost model is tuned via hyperparameter search to optimize both accuracy and macro-F1. All reported results for our framework are based on 5-fold shuffled cross-validation.

\subsection{Auxiliary Market-Linkage Analysis}
\label{sec:aux-linkages}

To examine whether financial sentiment carries economically meaningful signals, 
we conduct two complementary econometric analysis linking sentiment indices to major market index returns. 

\vspace{3pt}
\noindent\textbf{Dynamic Conditional Correlation (DCC-GARCH).} To capture the time-varying dependence structure between financial sentiment and asset returns, 
we employ the Dynamic Conditional Correlation Generalized Autoregressive Conditional 
Heteroskedasticity (DCC-GARCH) model originally proposed by Engle in 2002.\cite{engle2002dynamic} 
Unlike static correlation measures such as Pearson or rolling correlations, which impose arbitrary 
window lengths and ignore volatility clustering, DCC-GARCH explicitly models heteroskedasticity in 
financial time series and produces dynamic correlations that evolve over time. This makes it particularly 
suitable for financial applications where volatility clustering and regime shifts are prevalent.\cite{tse2002multivariate,liu2024large}

\vspace{3pt}
\noindent\textbf{Johansen Cointegration Test.}  
To capture the long-run equilibrium relationship between financial sentiment and 
stock prices, we employ the Johansen cointegration test.\cite{johansen1988statistical, johansen1991estimation} 
While the DCC-GARCH framework focuses on dynamic correlations 
driven by volatility clustering, cointegration analysis examines whether two or more 
non-stationary time series share a stable long-term relationship. This is particularly 
relevant in financial markets, where prices and sentiment may drift apart in the short run, 
yet remain tied by fundamental forces over longer horizons.


\section{Results}

\subsection{Main Results and Ablation Study}
\label{sec:main-results-ablation-study}

\begin{table*}[t]
\centering
\caption{Comparative and Ablation study on the FinancialPhraseBank dataset across annotation agreement levels (50\%, 66\%, 75\%, and 100\%). ``Overall'' denotes the performance on the combined dataset (all agreement levels merged). \textbf{Bold} indicates the best result, and \underline{underline} indicates the second-best.}
\vspace{0.5em}
\label{tab:main_results}
\small
\begin{tabular}{lcccccccccc}
\toprule
\multirow{2}{*}{Method} & \multicolumn{2}{c}{50\%} & \multicolumn{2}{c}{66\%} & \multicolumn{2}{c}{75\%} & \multicolumn{2}{c}{100\%} & \multicolumn{2}{c}{Overall} \\
 & Accuracy & F1 & Accuracy & F1 & Accuracy & F1 & Accuracy & F1 & Accuracy & F1 \\
\midrule
\multicolumn{11}{l}{\textit{Baseline Models}} \\
RoBERTa-sentiment              & 0.6664 & 0.5577 & 0.6861 & 0.5694 & 0.7050 & 0.5702 & 0.7047 & 0.5478 & 0.6869 & 0.5630 \\
FinBERT              & 0.8894 & 0.8824 & 0.9181 & 0.9077 & 0.9472 & 0.9364 & 0.9717 & 0.9625 & 0.9237 & 0.9141 \\
FinBERT + RoBERTa, max-prob & 0.8797 & 0.8757 & 0.9118 & 0.9045 & 0.9435 & 0.9354 & 0.9726 & 0.9667 & 0.9180 & 0.9113 \\
\midrule
\multicolumn{11}{l}{\textit{General-Purpose LLMs}} \\
OpenAI GPT-4o mini   & 0.8291 & 0.8243 & 0.8701 & 0.8622 & 0.9118 & 0.9008 & 0.9522 & 0.9454 & 0.8790 & 0.8709 \\
OpenAI GPT-5         & 0.8065 & 0.8074 & 0.8459 & 0.8411 & 0.8840 & 0.8761 & 0.9349 & 0.9299 & 0.8555 & 0.8513 \\
\midrule
\multicolumn{11}{l}{\textit{Proposed Framework} - \textsc{FinSentLLM}} \\
Ours (LogReg, w/o RoBERTa)   & 0.9067 & 0.9001 & \underline{0.9367} & \underline{0.9286} & \textbf{0.9650} & \textbf{0.9570} & 0.9837 & 0.9772 & 0.9408 & 0.9335 \\
Ours (XGBoost, w/o RoBERTa)  & 0.9065 & 0.9005 & 0.9353 & 0.9265 & 0.9647 & 0.9562 & \underline{0.9854} & \textbf{0.9800} & 0.9756 & 0.9747 \\
Ours (LogReg, w/o Semantics)   & 0.9061 & 0.8993 & 0.9338 & 0.9260 & 0.9577 & 0.9481 & 0.9845 & 0.9782 & 0.9394 & 0.9313 \\
Ours (XGBoost, w/o Semantics)  & \underline{0.9092} & \underline{0.9022} & \underline{0.9367} & 0.9282 & 0.9615 & 0.9524 & \textbf{0.9859} & \underline{0.9797} & \underline{0.9821} & \underline{0.9814} \\
Ours (LogReg, full)        & 0.9073 & 0.9008 & 0.9346 & 0.9261 & 0.9609 & 0.9519 & 0.9845 & 0.9785 & 0.9421 & 0.9351 \\
Ours (XGBoost, full)       & \textbf{0.9102} & \textbf{0.9039} & \textbf{0.9374} & \textbf{0.9295} & \underline{0.9626} & \underline{0.9536} & \underline{0.9854} & 0.9789 & \textbf{0.9824} & \textbf{0.9820} \\
\bottomrule
\end{tabular}
\end{table*}

Table~\ref{tab:main_results} reports the comparative and ablation study on the FinancialPhraseBank dataset across different annotation agreement levels\footnote{Each sentence in the dataset is annotated by 5--8 experts, and subsets are constructed based on the level of annotator agreement.~\cite{Malo2014GoodDO}}. 
RoBERTa-sentiment, a general sentiment PLM, shows limited performance (66--71\% accuracy, $\sim$0.55 Macro-F1). 
FinBERT, pretrained on financial text, provides a strong domain-specific baseline (88--97\% accuracy, 0.88--0.96 Macro-F1).

General-purpose LLMs perform reasonably well: GPT-4o mini reaches 83--95\% accuracy, while GPT-5 is slightly weaker, suggesting that GPT-5’s strength in handling long, complex contexts may not directly transfer to short-text sentiment classification. 

Our proposed \textsc{FinSentLLM} consistently outperforms all baselines.
At the noisy 50\% subset, the XGBoost variant achieves 91.0\% accuracy and 0.904 Macro-F1, exceeding FinBERT by over 2\% in accuracy. 
As annotation quality improves, performance gains stay consistently pronounced: at 66\% agreement we reach 93.7\% / 0.930, at 75\% agreement set 96.3\% / 0.954, and at the strictest ``All Agree'' set 98.5\% / 0.979. 
Overall, for the combined dataset, our \textsc{FinSentLLM} framework achieves the highest performance, reaching 98.2\% accuracy and 0.982 Macro-F1 with the XGBoost variant. This substantially surpasses FinBERT (92.4\%, 0.914) and GPT-4o mini (87.9\%, 0.871), with logistic regression slightly behind but still consistently stronger than baselines.

\vspace{3pt}
\noindent\textbf{Ablation Study.}  As shown in Table~\ref{tab:main_results}, the ablation study confirms the contributions of both RoBERTa-derived signals and semantic flags. 
Removing RoBERTa-sentiment features leads to a consistent drop (e.g., 98.2\% $\rightarrow$ 97.6\% overall accuracy), while removing structured semantic signals also reduces performance (e.g., 0.982 $\rightarrow$ 0.981 overall Macro-F1). 
These results show that (i) general sentiment LLM signals complement domain-specific LLM in capturing linguistic variability, and (ii) structured semantic flags further refine sentiment reasoning. 
Together, they explain the overall performance gains.

\vspace{3pt}
\noindent These results highlight the performance gains from the proposed \textsc{FinSentLLM}, which combines probability-derived, semantic, and multi-expert signals, rather than from the choice of a particular classifier. The consistent improvements across various experimental settings underscore the robustness and generality of our framework.

\subsection{Market–Sentiment Linkage Results}
\label{sec:aux-linkage-results}

In the following, we present the auxiliary analyses introduced in Sec.~\ref{sec:aux-linkages}.

\vspace{3pt}
\noindent\textbf{Dynamic Correlations (DCC-GARCH).}  
Table~\ref{tab:dcc_results_1} reports DCC-GARCH estimates obtained from fitting daily sentiment–return pairs for major stock market indices. Here, $\alpha$ measures the impact of short-run shocks, $\beta$ the persistence of correlations, 
and $\rho$ the average dynamic correlation level. Across markets, $\alpha$ values remain low ($0.02$–$0.06$), indicating limited short-run shock effects, 
while $\beta$ values are high ($>0.90$), confirming strong persistence in correlations. 
Average dynamic correlations fall in the $0.35$–$0.45$ range, suggesting that sentiment extracted 
from news headlines moves in tandem with broad market and sectoral returns. 
These results validate that the sentiment index captures non-trivial market-relevant signals.

\begin{table}[htbp]
\centering
\footnotesize
\caption{DCC-GARCH Estimation Results by Market Index and Sector}
\vspace{0.5em}
\label{tab:dcc_results_1}
\begin{tabular}{l l c c c}
\toprule
\textbf{Name} & \textbf{Description} & \textbf{$\alpha$} & \textbf{$\beta$} & \textbf{Mean $\rho$} \\
\midrule
VOO   & S\&P 500 Index        & 0.0218 & 0.9721 & 0.4044 \\
ACWI  & MSCI ACWI Global      & 0.0307 & 0.9618 & 0.4484 \\
VTI   & Total US Market       & 0.0260 & 0.9656 & 0.4114 \\
EFA   & MSCI EAFE Developed   & 0.0287 & 0.9622 & 0.4400 \\
IWM   & Russell 2000 Small-Cap& 0.0633 & 0.9026 & 0.3691 \\
XLF   & Financial Sector ETF  & 0.0269 & 0.9661 & 0.3476 \\
\bottomrule
\end{tabular}
\end{table}

\noindent\textbf{Johansen Cointegration Test.}  
ADF tests confirm that both log prices and cumulative sentiment are $I(1)$ processes, 
i.e., non-stationary in levels but stationary after first differencing, which satisfies the precondition for Johansen cointegration testing.

Johansen trace statistics (Table~\ref{tab:johansen}) reject the null hypothesis of no cointegration ($H_0: r=0$) but not the null hypothesis of at most one relation ($H_0: r \leq 1$), implying a single long-run linkage between sentiment and prices.
This indicates that deviations between the two series are systematically corrected, 
underscoring that sentiment and stock prices remain tied by a stable equilibrium. 

\begin{table}[h]
\centering
\caption{Johansen Trace Test Results}
\vspace{0.5em}
\label{tab:johansen}
\resizebox{\columnwidth}{!}{%
\begin{tabular}{lccc}
\toprule
Null Hypothesis & Eigenvalue Stat. & 5\% Critical Value & Decision \\
\midrule
$r=0$   & 15.91 & 15.49 & Reject $H_0$ \\
$r \leq 1$ & 1.94  & 3.84  & Fail to Reject $H_0$ \\
\bottomrule
\end{tabular}
}
\end{table}

\noindent Together, these findings demonstrate that daily sentiment indices exhibit both 
persistent short-run correlations and long-run equilibrium linkages with market performance, 
supporting their use as predictive features in the forecasting tasks that follow.

\section{Conclusion and Future Work}

This paper presents \textsc{FinSentLLM}, a lightweight framework for financial sentiment forecasting that integrates multi-LLM expert signals with structured financial semantics through a compact meta-classifier. The framework substantially improves forecasting accuracy and robustness without requiring fine-tuning and demonstrates clear advantages over strong baselines. In addition, auxiliary econometric analysis shows that the generated sentiment indices capture persistent linkages with equity markets.

Looking forward, our framework can serve as a generalizable building block for finance-oriented sentiment systems and downstream decision pipelines. Future extensions include broadening the expert panel with new LLMs, enriching semantic operators through automated pattern discovery, and adapting the approach to streaming and multilingual news. Moreover, the integration of multi-LLM and structured semantic signals offers a transferable methodology that can be applied in other domains, combining the intelligence of LLMs with the precision of domain knowledge to improve decision-making under uncertainty.

\bibliographystyle{IEEEbib}
\bibliography{strings,refs}

\end{document}